# The ALMA Development Program: Roadmap to 2030

John Carpenter (Joint ALMA Observatory), Daisuke
Iono (National Astronomical Observatory of Japan),
Francisca Kemper (European Southern Observatory),
and
Al Wootten (National Radio Astronomy Observatory)

## 1 Introduction

The Atacama Large Millimeter/submillimeter Array (ALMA) is the premier telescope for sensitive, high-resolution observations at millimeter and submillimeter wavelengths. The array consists of fifty 12-m diameter antennas that can be reconfigured to baselines as long as 16 km, twelve 7-m antennas that sample the short visibility spacings, and four 12-m antennas that provide total power capabilities for spectral line and continuum observations. Located in the Atacama desert in northern Chile at an elevation of 5000 m on the Chajnantour plateau, the ALMA site provides excellent observing conditions with low precipitable water vapor. The large number of antennas, the high-altitude site, and excellent receivers with low-noise performance provide an extremely sensitive, flexible instrument for submillimeter imaging.

The broad theme of ALMA science is to understand our cosmic origins, from the formation of galaxies to the formation of planets. Figure 1 illustrates the spectacular science return from ALMA across a breadth of fields. Galaxies have been detected at redshift as high as $z = 9.11$, which implies the onset star formation began as early as 250 Myr after the Big Bang (Hashimoto et al., 2018). Very Long Baseline Interferometry (VLBI) with ALMA and the Event Horizon Telescope has produced the first image of a black hole (Event Horizon Telescope Collaboration et al., 2019). High resolution ALMA images are revealing the formation of young stars (see, e.g., Matsushita et al., 2019). The iconic image of the cirumstellar disk associated with the young star HL Tau revealed a nested network of concentric rings that suggest the onset of planet formation is much earlier than previously thought (ALMA Partnership et al., 2015).

In addition to providing funds for the annual operating budget, the ALMA partners[1] contribute funds for new ALMA developments. This development fund is designed to enhance ALMA as the start-of-the-art and world leading facility for millimeter/submillimeter astronomy by promoting hardware, software, and infrastructure improvements for ALMA. The development program relies on a close collaboration between ALMA and the broader community. ALMA provides high-level guidance on the overall priorities for the development program and provides funding support for innovative ideas from the community to achieve these goals.

As ALMA approaches completion of its initially envisaged capabilities, the original fundamental science goals of ALMA have essentially been demonstrated. Looking to the future, the ALMA Board tasked the ALMA Science Advisory Committee (ASAC) to recommend future developments that ALMA should consider implementing by the year 2030. A working group prioritized those recommendations into a strategic ALMA2030 Development Roadmap, which was then endorsed by the ALMA Board in 2017 (Carpenter et al., 2019). The Roadmap identifies three new science goals to motivate technical developments over the next decade under three basic themes: Origins of Galaxies, Origins of Chemical Complexity, and Origins of Planets (Figure 2). Achieving these science goals requires major new advances in ALMA capabilities, and in particular in improving the instantaneous bandwidth of the receivers, the associated electronics, and the correlator.

---

[1]ALMA is a partnership between the European Southern Observatory (ESO), the National Science Foundation (NSF) of the United States and the National Institutes of Natural Sciences (NINS) of Japan in collaboration with the Republic of Chile. ALMA is funded by ESO in representation of its member states, by NSF in collaboration with the National Research Council (NRC) of Canada and the National Science Council (NSC) of Taiwan, and by NINS in collaboration with the Academia Sinica (AS) in Taiwan, and the Korea Astronomy and Space Science Institute (KASI) of South Korea.

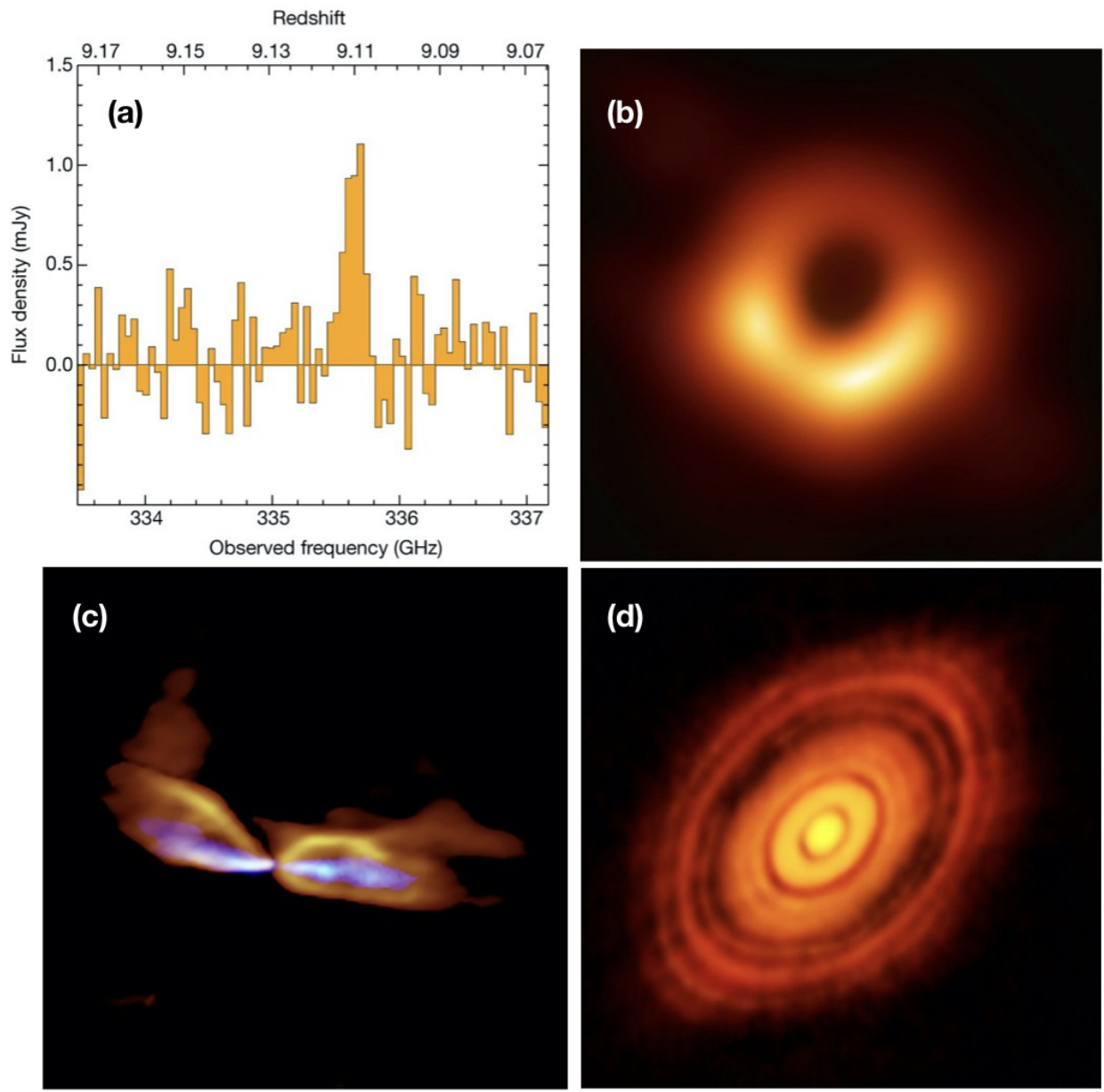

Figure 1: A small sample of ALMA observations. (a) Detection of [O iii] toward a galaxy at redshift $z = 9.11$ (Hashimoto et al., 2018). (b) Image of the super massive blackhole at the center of the galaxy M87 produced by ALMA and the Event Horizon Telescope (Event Horizon Telescope Collaboration et al., 2019). (c) Images of the low velocity outflow (orange) and the high velocity jet (blue) from a protostar in Orion (Matsushita et al., 2019). (d) ALMA continuum image of the protoplanetary disk surrounding the young star HL Tau (ALMA Partnership et al., 2015).

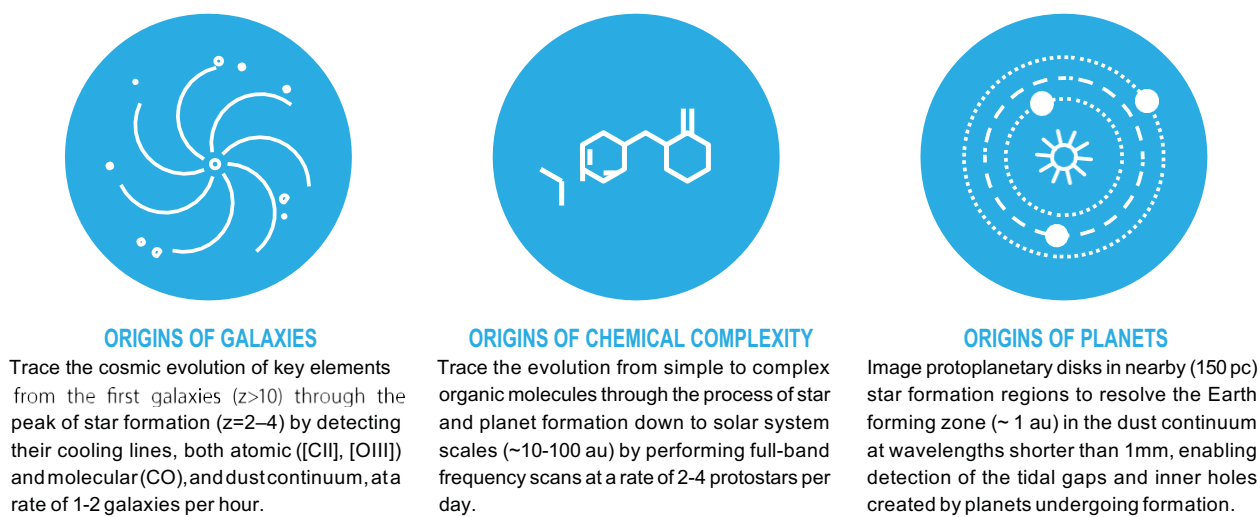

Figure 2: Science goals for the ALMA 2030 Development Roadmap.

In this article, we provide an overview of the ALMA development program. We begin by summarizing the organization of the regional development efforts. We then describe the current characteristics and performance of the ALMA receivers. We conclude with a summary of the ongoing progress toward achieving the Roadmap and the development workshops in the upcoming year that are being organized by ALMA.

## 2  Overview of the ALMA Development Program

NAOJ, ESO, and NRAO administer their respective regional contribution to the ALMA Development Program. Each executive holds development workshops to gather input from the community on future developments, release call for proposals for development proposals, oversee the selection of the proposals, and administer the selected programs.

ALMA has two categories of development programs: *studies* and *projects*. Studies are funded by and managed at the discretion of the regional executives. Studies are designed and intended to allow an initial exploration of concepts that may be of interest to ALMA development in the near or long terms. Projects are larger-scale programs that provide specific deliverables (either hardware or software) to ALMA. Because of the unique circumstances in each region, the detailed implementation of studies and projects vary between the regions. The following subsections describes the local execution of the regional development programs and ongoing activities.

## 2.1 East Asia

The East Asia (EA) ALMA development program is driven by community input in the annual EA Development Workshops as well as the development priorities recommended by the EA ALMA Scientific Advisory Committee (EASAC) and the ASAC. NAOJ leads the development activities in EA ALMA, and collaborates with its partner institutes in Taiwan (ASIAA) and Korea (KASI). EA ALMA has organized seven development workshops since 2011, with the main aim of discussing the technical feasibility and science demands from the EA ALMA science community.

Besides collaborating with ESO on the development of the Band 2 receiver (see Section 3.2), EA ALMA currently leads two development projects:

- The Band 1 receiver, which will provide frequency coverage between 35 and 50 GHz. Section 3.1 discusses the Band 1 receiver development.

- The Atacama Compact Array (ACA) spectrometer for the Total Power Array. The ACA spectrometer development is a project led by KASI in collaboration with NAOJ. This new GPU based spectrometer improves the linearity, dynamic range and spectral response over the existing ACA correlator. The proposed simple architecture and the software nature allows seamless implementation of new capabilities in the future. Science observations with the total power spectrometer are expected to begin in October 2021.

In addition to the two development projects, the EA ALMA leads four important development studies which can significantly enhance the future capabilities of ALMA. They are (1) a state- of-the-art high-Jc SIS junction studies for wide IF/RF, (2) an Artificial Calibration Source for polarization calibration, (3) a stable LO reference source for future longer baselines, and (4) a multi-beam receiver using an on-chip integrated circuit. All of the development studies are aligned with the recommendations given in the Roadmap.

## 2.2 Europe

ESO coordinates and funds the European development program on behalf of the ESO member states. Development studies are organized in 3-year cycles, in which roughly 5-10 studies are approved for a budget not exceeding 50 kEUR/yr per study over at most a 3-year period. The most recent call for proposals for development studies was issued at the end of May 2019 with a deadline in September 2019. In this call, proposal teams were encouraged to aim for alignment with the Roadmap. The call was accompanied by a European ALMA Development workshop, held at ESO Garching from 3-5 June 2019. The workshop was open to participants from all ALMA regions, and the full breadth of the ALMA development program was discussed[2]. A list of approved European development studies is maintained online[3].

Current European development programs encompasses a small number of recently finished and on-going projects:

- The Integrated Alarm System is a software development project run by ESO that allows the ALMA telescope operators to monitor several systems at once, including weather stations and antenna status monitors. The software has been running at ALMA since late August 2019.

---

[2] https://www.eso.org/sci/meetings/2019/ALMADevel2019.html
[3] https://www.eso.org/sci/facilities/alma/development-studies.html

- The Band 2 receiver development is a project coordinated at ESO with contributions from institutes worldwide, including many across Europe, Japan, Chile, and the US. Section 3.2 discusses the current status of the Band 2 receiver development.

- The Additional Representative Images for Legacy project (ARI-L) is a development project led by the Istituto Nazionale di Astrofisica (INAF) in Bologna, which also hosts the Italian ARC-node. The project formally began in June 2019. The goal is to reprocess eligible data from Cycles 2-4 with the current data reduction and imaging pipelines in order to produce quick-look images and spectral datacubes for >70% of the data. These data products will be ingested into the ALMA Science Archive to facilitate archival science. The project will also provide the pipeline-calibrated measurement sets.

## 2.3 North America

The North America (NA) development program is open to the NA ALMA Operations Partnership, which is defined as the community of astronomers and scientists in related fields from North American ALMA partner countries. Since 2011, thirty-nine study proposals and eleven project proposals from the NA community have been funded. Reports from these studies are available at the NRAO website[4].

A Call for Study proposals in North America is issued on a yearly basis, with the most recent call[5] issued in December 2019 for FY2021. Studies are generally funded for one year up to $250,000 USD per individual award for FY2021. Priority is currently given to those studies which align with the Roadmap. Study topics of particular interest to the NA ALMA Partnership include larger bandwidths and improved receiver sensitivity, longer baselines, increasing wide field mapping speed, phased array feeds, and improvements to the data archive.

An independent review panel is established with consent of the National Science Foundation (NSF) to evaluate and rank the proposals. The ranked list then receives consent from NRAO and NSF and is incorporated into a recommendation to the NA ALMA Executive. The NA ALMA Executive has funding authority, and responsibility, for executing the NA ALMA Development Studies plan.

The active NA projects are as follows:

- Expansion of the Central Local Oscillator Article (CLOA) to Five Subarrays: This project, led by NRAO, procured and tested all the required modules and equipment to complete Photonic LO subarray five. The complete chain was installed, tested, and commissioned at the AOS Technical Building. The completed system was integrated into the current software control system. The completion of this project is pending final test results and acceptance of the final report by ALMA.

- Band 3 Cold Cartridge Assembly (CCA) Magnet and Heater Installation for Deflux Operations: This project, led by the NRC-Herzberg Astronomy and Astrophysics (NRC-HAA) Research Center, is modifying the Band 3 CCA to add a heater element in order to reduce observed azimuth-dependent total power variations. The heater solution was successfully tested at NRC-HAA and underwent verification testing by the JAO. The design has now been finalized and NRC-HAA is now building the initial heater kits for delivery in 2020. Integration into each Band 3 CCA will continue over the successive three years.

---

[4] https://science.nrao.edu/facilities/alma/alma-develop-old-022217/alma-develop-history
[5] https://science.nrao.edu/facilities/alma/science_sustainability/cycle8-cfs

- ALMA Phasing System Phase 2 (APP2): This project, led by MIT Haystack, will further improve VLBI capabilities and performance for ALMA. Major components include enabling spectral line VLBI, extending the frequency range of phasing to Bands 1–7, improving the calibration mechanism to allow observations on weaker sources, the introduction of a single-dish VLBI mode, and a pulsar mode. On-sky testing has been conducted during multiple VLBI campaigns at ALMA in coordination with other observatories. Passive phasing and pulsar mode will be offered in Cycle 8.

## 3 ALMA receivers

The ALMA front end can accommodate up to 10 receiver bands covering most of the wavelength range from 8.5 to 0.32 mm (35-950 GHz). Each band is designed to cover a tuning range which is approximately tailored to the atmospheric transmission windows. These windows and the tuning ranges are indicated in Figure 3.

The ALMA receivers in each antenna are located in a single front-end assembly. The front-end assembly consists of a large cryostat containing the receiver cold cartridge assembly for each band (including the Superconductor-Insulator-Superconductor (SIS) mixers and Local Oscillator (LO) injection) and the Intermediate Frequency (IF) and LO room-temperature electronics of each band. The cryostat is kept at a temperature of 4 K through a closed cycle cooling system. Each receiver cartridge contains two complete receiving systems sensitive to orthogonal linear polarizations. The designs of the mixers, optics, LO injection scheme, and polarization splitting vary from band to band, depending on the optimum technology available at the different frequencies.

Table 1: The ALMA receivers

| Band | RF (GHz) | IF (GHz) | Mixer type |
|---|---|---|---|
| 1 | 35-50 | 4-12 | SSB |
| 2 | 70-95 | 4-12 | 2SB |
| 3 | 84-116 | 4-8 | 2SB |
| 4 | 125-163 | 4-8 | 2SB |
| 5 | 158-311 | 4-8 | 2SB |
| 6 | 211-275 | 4.5-10 | 2SB |
| 7 | 275-373 | 4-8 | 2SB |
| 8 | 385-500 | 4-8 | 2SB |
| 9 | 602-720 | 4-12 | DSB |
| 10 | 787-950 | 4-12 | DSB |

Table 1 summarizes the basic characteristics of the bands, include the Radio Frequency (RF) range, the Intermediate Frequency (IF), and mixer type (2SB: dual sideband receiver; SSB: Single sideband; DSB: double sideband). Figure 4 shows the measured receiver noise temperatures for the currently available bands. For Bands 3–7, the receiver temperatures are approximately $4h\nu/k$ and are approaching quantum-limited performance. Detailed receiver characteristics are available in ALMA Technical Handbook available on the ALMA Science Portal[6]. The following subsections highlight the anticipated characteristics of Band 1 and Band 2, which are the next ALMA receivers to be installed.

[6] https://almascience.org/documents-and-tools/latest/alma-technical-handbook

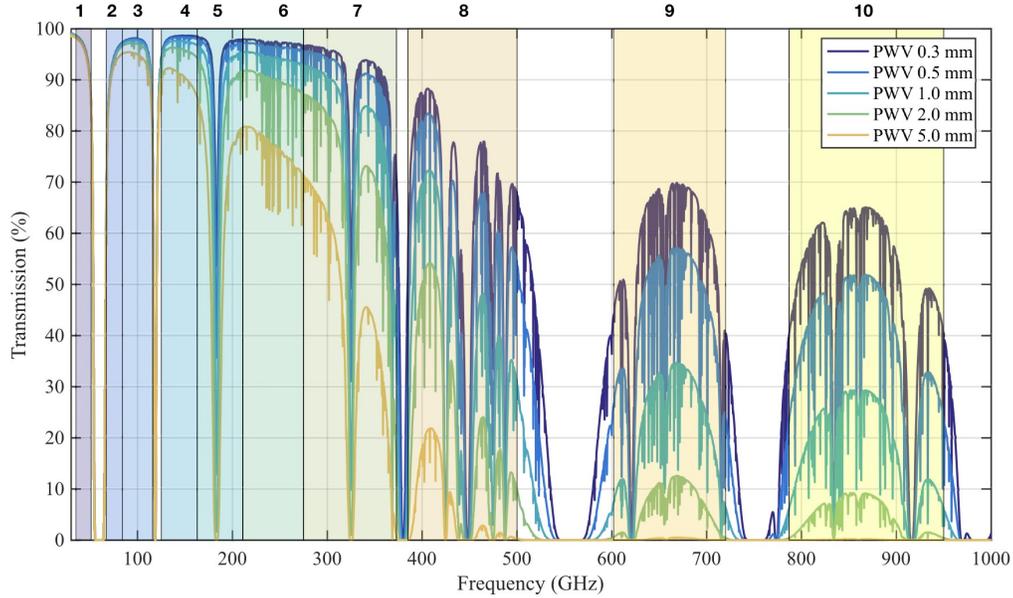

Figure 3: The ten ALMA receiver bands along with atmospheric transmission. The receiver coverage is shown shaded, superimposed on a zenith atmospheric transparency plot at the ALMA Array Operation Site (AOS) for 0.3, 0.5, 1.0, 2.0 and 5.0 mm of precipitable water vapor (PWV). The ALMA band numbers are indicated at the top of the figure in bold.

## 3.1 Band 1

The Band 1 receiver covers the lowest frequency band of ALMA from 35 to 50 GHz, and it is a development project led by ASIAA in collaboration with NAOJ, NRAO, University of Chile and NRC-HAA. As described in Di Francesco et al. (2013), the range of Band 1 science is very broad, from solar studies, nearby stars, protoplanetary disks, and molecules in nearby molecular clouds to distant galaxy clusters and the re-ionization edge of the Universe.

One of the primary science drivers for Band 1 is to measure the grain properties in protoplanetary disks. The formation of planets require the agglomeration of dust grains into larger particles. The long wavelength observations provided by Band 1 can probe larger particles than the higher-frequency ALMA bands, and the optical depth of the continuum emission in the disk will also be reduced. By observing a diverse sample of disks in Band 1, it will be possible to explore where and when in the disk that dust agglomeration takes place. A second primary science driver for Band 1 is to identify galaxies in the epoch of reionization. By targeting known candidate high-redshift galaxies, Band 1 will measure the spectroscopic redshift, especially using the $J$=3-2 transition of carbon monoxide, which is redshifted into the Band 1 frequency range for redshifts of $6 \lesssim z \lesssim 9$.

The technology used for Band 1 is dual-polarization, SSB heterodyne receiver covering the specified frequency range with an IF band of 4-12 GHz. Band 1 integration at the ALMA Operations Support Facility (OSF) will start in the early part of 2020, with a target completion to outfit all 66 antennas with Band 1 receivers by the end of 2022. Figure 5 shows the measured noise performance of a prototype Band 1 receiver.

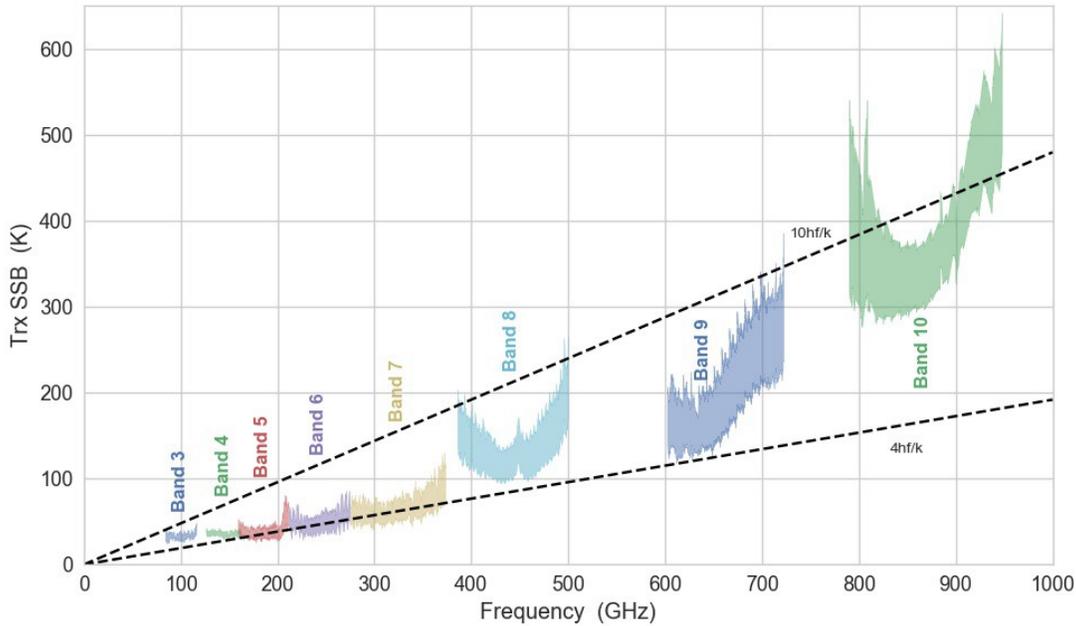

Figure 4: Receiver noise temperature for ALMA receivers Band 3-10. The shaded region encompasses 75% of the receivers about the median receiver temperature. The noise temperature shown for Band 9 and 10 are twice the DSB temperatures. Figure from Carpenter et al. (2019).

## 3.2 Band 2

The Band 2 receiver project is in the prototyping stage with the pre-production of six units. The goal is to cover a RF bandwidth of 67–116 GHz, making use of the full atmospheric window that includes ALMA Band 2 and Band 3. For the optical components, a design from NAOJ has been selected. The selection of the Low Noise Amplifiers (LNAs) is ongoing, with prototypes from the University of Manchester and the Low Noise Factory currently being tested (Yagoubov et al., 2019). Also, a prototype LNA from the Cahill Radio Astronomy Laboratory CRAL (NRAO) is expected to be delivered in early 2020. Figure 6 shows the performance of the prototype receiver. The LNA downselection is expected to be completed by mid-2020. The production of the required 73 units is expected to be complete in 2023, and integration and commissioning at OSF will take place over the period 2022–2024.

## 4 Toward ALMA 2030

The Roadmap identifies three main areas to drive the developments over the next decade.

- The top development priority, based on scientific merit and technical feasibility, is to broaden the receiver IF bandwidth by at least a factor of two and to upgrade the associated electronics and correlator to process the entire bandwidth.

- Identify and add the functionality needed for the community to mine the ALMA archive efficiently, especially in view of the increased receiver bandwidth envisioned in the Roadmap.

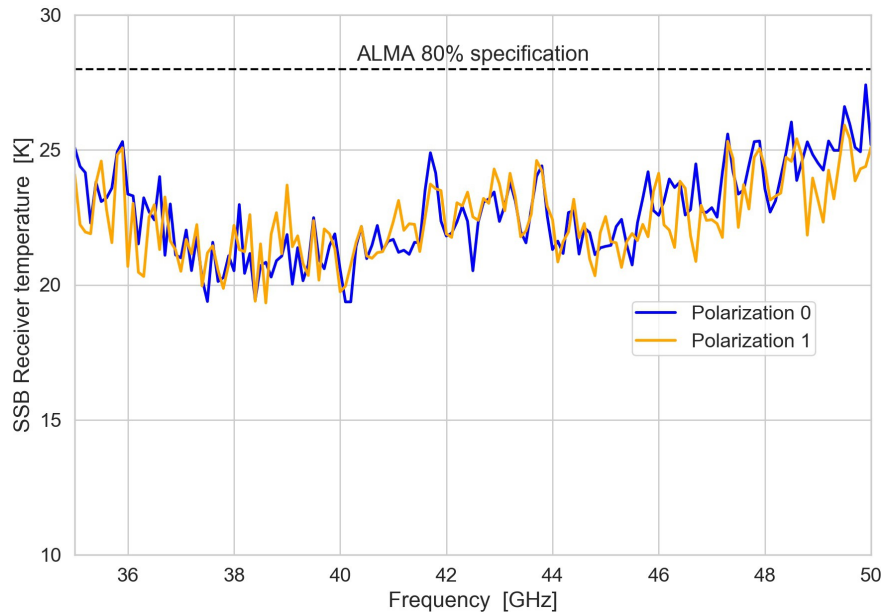

Figure 5: Single-sideband receiver temperature for a Band 1 prototype receiver. The solid curves indicates the measured noise temperature for the two polarizations. The dashed curve indicates the technical specification of the receiver, which is that the noise temperature must be less than 28 K in 80% of the band. Plot courtesy of the Band 1 development team.

- Continued exploration of development paths which have potentially large impacts on ALMA science, but for which the science case and technical feasibility require further investigation. These include (i) extending the maximum baseline length by a factor of 2-3 to image the terrestrial planet forming zone in nearby protoplanetary disks; (ii) develop focal plane arrays to significantly increase ALMA's wide-field mapping speed; and (iii) increasing the number of 12-m antennas, to benefit all science programs by improving the sensitivity and image fidelity.

The ALMA Development Program is already supporting studies and projects toward all of these goals, and encourages the community to continue submitting proposals. This section highlights two development paths in particular that have made significant progress: increasing the instantaneous IF bandwidth in the next-generation ALMA receivers and extending the maximum baseline length.

## 4.1 Next-generation ALMA receivers

The current ALMA digital processing signal and correlator handle 16 GHz of bandwidth (8 GHz per polarization). However, the 8-GHz instantaneous frequency coverage per polarization covers a small fraction of the atmospheric windows. Expanding the throughput by a factor of two or more will reduce the time to conduct blind redshift surveys by the corresponding factor. Wide bandwidths can also observe multiple lines in a galaxy simultaneously to secure redshifts and measure the physical conditions of the interstellar medium. Chemical spectral scans at high spectral resolution, commonly needed for galactic sources, will be an order of magnitude faster when combined with an upgraded correlator (see Figure 7). Increased bandwidth will also improve the continuum sensitivity.

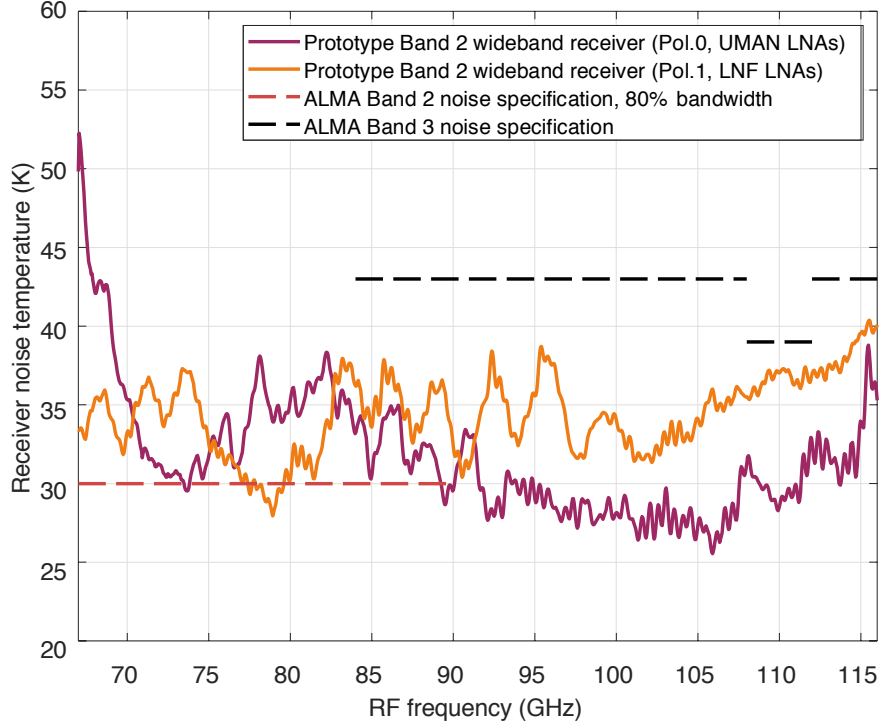

Figure 6: Measured receiver noise temperature for two polarisation channels of the prototype Band 2 receiver. The channel labeled 'Pol. 0' is based on an LNA from the University of Manchester (UoM), and that labeled 'Pol. 1' is based an LNA from the Low Noise Factory (LNF). Figure from Yagoubov et al. (2019).

Impressive progress toward wide IF bandwidth receivers is already underway. As shown in Section 3, bands 1 and 2 will have 8 GHz of IF bandwidth. Designs for even larger bandwidths are being explored. NAOJ is pushing the limits of RF and IF bandwidths of heterodyne receiver front ends based on high-Jc SIS mixers. The utilization of the high-Jc SIS junction offers wide RF and IF bandwidths thanks to their lower RC product. In addition, the wide ranges of Jc and junction sizes increase the design flexibility in the RF and IF circuits. These properties facilitate better IF power coupling from the SIS junction to the standard 50-Ohm IF circuit or better matching between SIS junctions and cryogenic low noise amplifier. The heterodyne module developed at NAOJ integrates the SIS mixer and cryogenic low-noise amplifier in a single block, omitting an isolator. This work has led to the demonstration of low-noise double sideband SIS mixers covering ALMA bands 7 and 8 and IF bandwidths from 3 to 22 GHz, as shown in Figure 8. NAOJ is studying optimum sideband separating mixer configuration with the wide IF bandwidth and feasibility of a high-speed analog-digital converter in order to implement them into the telescope as a full receiver system, and also plans to apply these technologies to other frequency bands; e.g., ALMA Band 10.

## 4.2 Baseline extension

An ongoing study led by ESO addresses the feasibility of extending the maximum baselines to as long as 32 km, or double the current baseline length. Figure 9 shows the pad positions within the ALMA concession and the Atacama Astronomical Park that are being considered for additional baselines. The technical feasibility of such long baselines, considering issues such as the impact of long fiber connections, delay tracking, and atmospheric calibration, is being evaluated using a testbed interferometer between telescopes at the Array Operations Site (AOS) and the site of the Observatory Support Facility (OSF). The tests have shown repeatably that synchronization, receiver locking and digital data transport can work reliably over the 30 km AOS-OSF fiber connections. Minor upgrades in software and firmware (of the digitizer clock) to increase the usable range of delays and delay rates have been successfully tested, and fringes in Bands 3, 5, and 7 were reliably obtained on a 23.4 km baseline in June 2019. So far all tests have been performed with a single baseline interferometer based on spare correlator and LO resources at the OSF, and future tests will extend this to using the AOS production correlator and LO resources and large arrays of antennas. During 2019, optimizations to delay dispatching in the correlator software have been made which will facilitate extended baselines in addition to improving reliability and lowering overheads for current long baseline observations. In the upcoming year, ALMA will develop a detailed science case for long baselines and designing an array configuration that makes use of all 66 antennas.

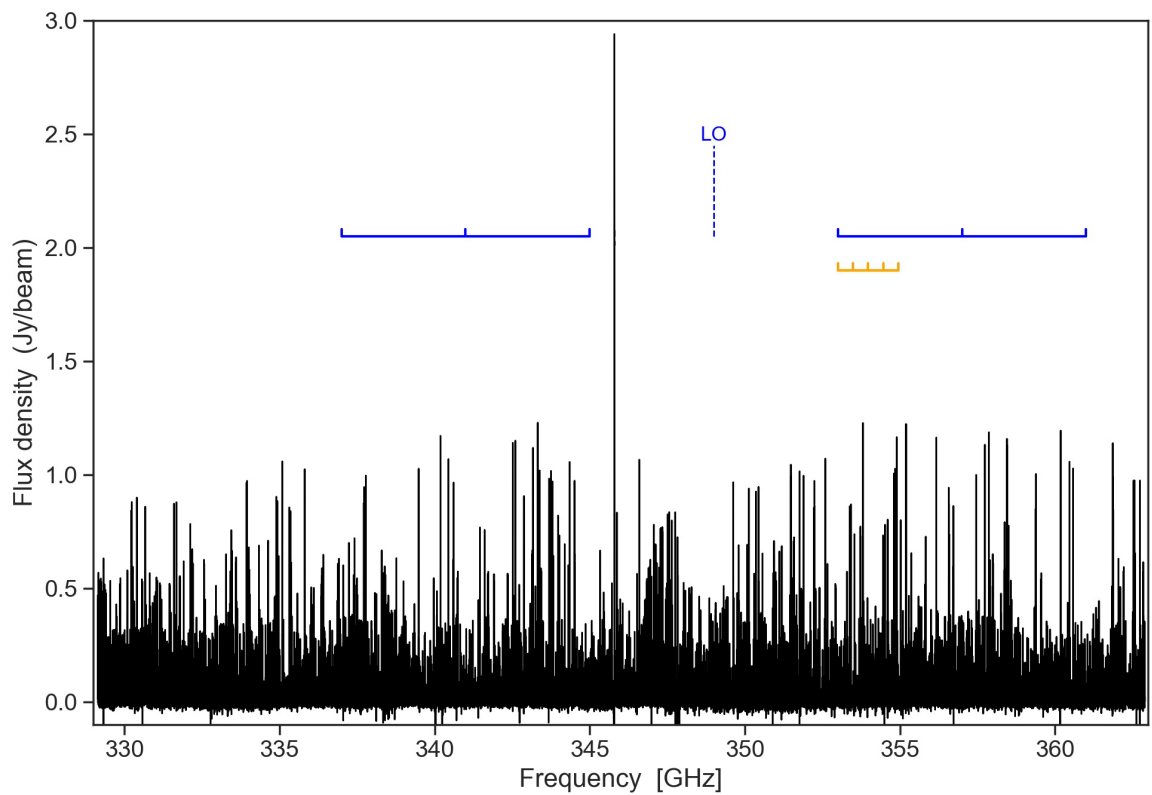

Figure 7: Spectral line survey of the protostellar binary IRAS 16293-2422 obtained using ALMA (Jørgensen et al., 2016). The observations required 18 individual tunings given the instantaneous spectral coverage of the current system at a spectral resolution of 0.2 km s$^{-1}$ (orange). Upgrading the receivers, electronics, and correlator to the *minimum* specifications of the Roadmap will make spectral surveys at least an order of magnitude faster.

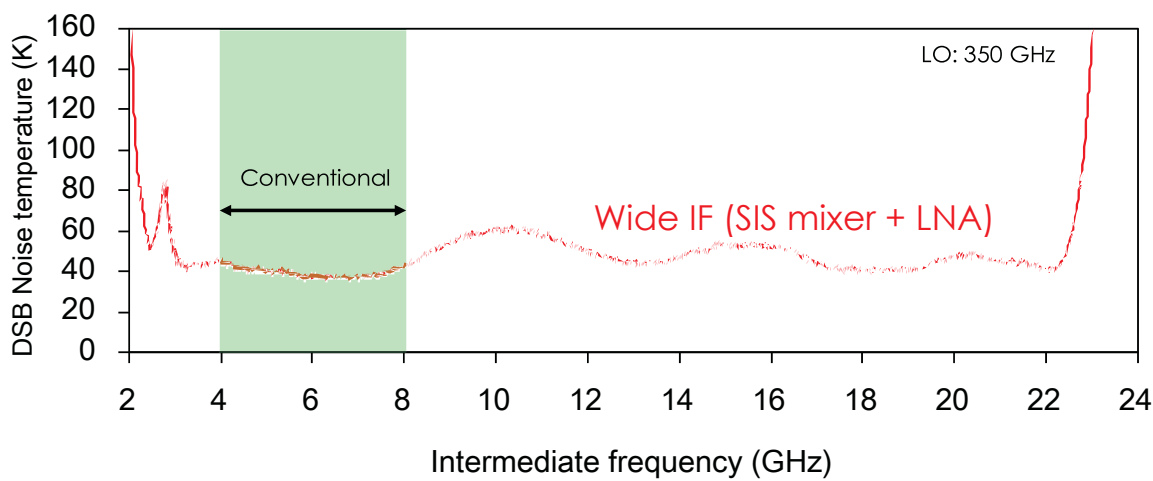

Figure 8: Double sideband noise temperature of an wide-IF SIS receiver at a local oscillator frequency of 350 GHz as a function of intermediate frequency. Figure from Kojma (2019).

## 4.3 ALMA 2030 Development Workshops in 2020

Community involvement is essential to perform the upgrades envisioned in the ALMA Development Roadmap. Over the next year, ALMA is hosting workshops that will help define the technical requirements and select the technologies needed for the upgrades. The workshops include:

- The ALMA2030 Vision: Design Considerations for the Next ALMA Correlator[7]
  The workshop will bring together experts on the ALMA system and modern digital correlator design in order to (1) discuss design requirements for the next generation correlator that enables the ALMA2030 vision; (2) share pros and cons of recent and currently under design correlator architectures; and (3) identify challenges for implementing and deploying a new ALMA correlator. The workshop will be held 11-13 February 2020 in Charlottesville, Virginia.

- The ALMA 2030 Vision: Design considerations for Digitizers, Backend and Data Transmission System[8]
  This meeting will bring together experts on the ALMA system and digitizer, backend and data transmission system technologies in order to (1) discuss the status of technology and performance prospects for the next decade for digitizers, backend and DTS; (2) identify the impact on these subsystems due to the increase of the instantaneous bandwidth by a factor $>2$; (3) discuss the most suitable location of the second generation ALMA correlators in relation with the status of DTS technologies; (4) identify the possibility to use off-the-shelf technologies for the implementation of the second generation of ALMA digitizers, backend and DTS; and (5) discuss possible system architectures to implement the multiplication of the IF bandwidth of ALMA by a factor $>2$. The workshop will be held 11-13 March 2020 in Mitaka, Japan.

- ALMA is also organizing a workshop on the latest in front-end technology to be held in Europe in late 2020. The details of the workshop will be announced soon.

ALMA strongly encourages the community to participate in the workshops. These workshops will help establish the specific requirements for the next generation receivers and correlators, and determine the technical directions needed to achieve the goals of the Roadmap.

---

[7] http://go.nrao.edu/NextALMACorrelator
[8] https://alma-intweb.mtk.nao.ac.jp/~diono/meetings/ALMA2030_Mitaka

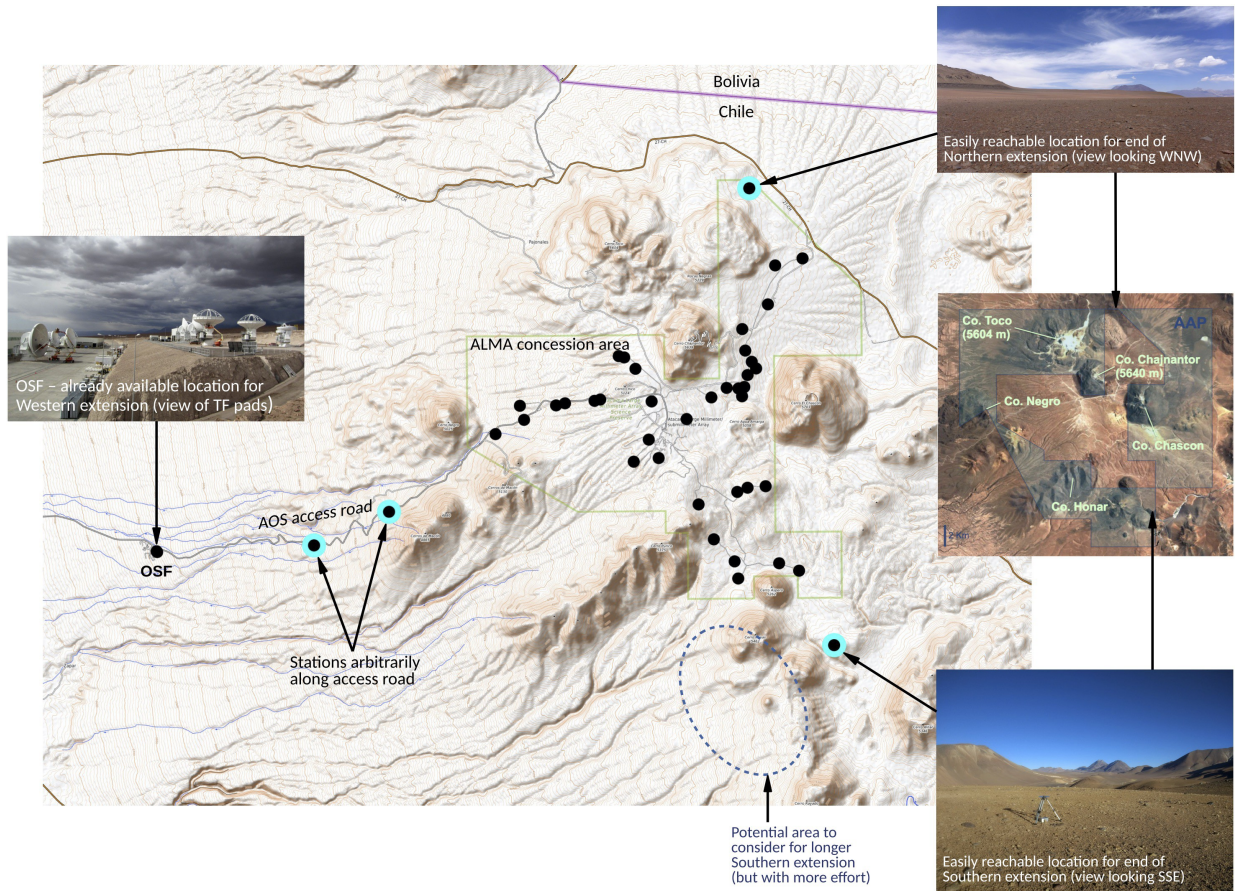

Figure 9: Baseline length extension by up to a factor of ~ 2 within the current operational model. On the map, the plain black points denote the most remote existing stations, including at the OSF. The black points highlighted in cyan denote viable new station locations. To the west of the existing array, stations can be added along the AOS access road between the OSF and the AOS. To the north, stations can be added up to 4 km further than now within the existing ALMA concession. To the south, stations can easily be added up to 4 km further than now within the Atacama Astronomical Park area. Technical feasibility is currently being established with an interferometer between array elements at the OSF and the AOS sites on a 23.4 km baseline and 30 km fiber paths.